\documentclass[letterpaper]{article} 
\usepackage{aaai2026}  
\usepackage{times}  
\usepackage{helvet}  
\usepackage{courier}  
\usepackage[hyphens]{url}  
\usepackage{graphicx} 
\usepackage{natbib}  
\usepackage{caption} 
\usepackage{algorithm}
\usepackage{algorithmic}
\usepackage{amssymb}
\usepackage{amsmath}
\usepackage{booktabs} 
\usepackage{tabularx}
\usepackage{multirow} 
\usepackage{subcaption}
\usepackage{newfloat}
\usepackage{listings}
\DeclareCaptionStyle{ruled}{labelfont=normalfont,labelsep=colon,strut=off} 
\floatstyle{ruled}
\newfloat{listing}{tb}{lst}{}
\floatname{listing}{Listing}
\newcommand{\methodname}{HW-GNN}
\title{HW-GNN: Homophily-Aware Gaussian-Window Constrained Graph Spectral Network for Social Network Bot Detection}
\author{
    Zida Liu$^{*}$,
    Jun Gao$^{*}$,
    Zhang Ji$^{\dagger}$,
    Li Zhao$^{\ddagger}$
}
\affiliations{
    $^{*}$Key Laboratory of High Confidence Software Technologies, CS, Peking University, China\\
    $^{\dagger}$Nanjing University of Aeronautics and Astronautics, China\\
    $^{\ddagger}$Zhejiang Lab, Zhejiang, China\\
    zdliu25@stu.pku.edu.cn, gaojun@pku.edu.cn, zhangji77@gmail.com, lzjoey@gmail.com
}

\nocopyright  
\begin{document}

\maketitle



\begin{abstract}
Social bots are increasingly polluting online platforms by spreading misinformation and engaging in coordinated manipulation, posing severe threats to cybersecurity. Graph Neural Networks (GNNs) have become mainstream for social bot detection due to their ability to integrate structural and attribute features, with spectral-based approaches demonstrating particular efficacy due to discriminative patterns in the spectral domain. However, current spectral GNN methods face two limitations: (1) their broad-spectrum fitting mechanisms degrade the focus on bot-specific spectral features, and (2) certain domain knowledge valuable for bot detection, \emph{e.g.}, low homophily correlates with high-frequency features, has not been fully incorporated into existing methods.

To address these challenges, we propose \methodname{}, a novel homophily-aware graph spectral network with Gaussian window constraints. Our framework introduces two key innovations: (i) a Gaussian-window constrained spectral network that employs learnable Gaussian windows to highlight bot-related spectral features, and (ii) a homophily-aware adaptation mechanism that injects domain knowledge between homophily ratios and frequency features into the Gaussian window optimization process. Through extensive experimentation on multiple benchmark datasets, we demonstrate that \methodname{} achieves state-of-the-art bot detection performance, outperforming existing methods with an average improvement of 4.3\% in F1-score, while exhibiting strong plug-in compatibility with existing spectral GNNs.
\end{abstract}

\section{Introduction}

Social bots have become a pervasive threat in online social networks~\cite{abulaish2020socialbots}, deceiving users by disseminating false information, amplifying harmful content, and coordinating manipulative campaigns~\cite{domalewska2021disinformation,ferrara2017disinformation} that can influence elections and undermine trust in digital platforms~\cite{hajli2022social}. The detection of social bots is particularly challenging due to their rapidly evolving sophistication. Modern social bots are capable of closely mimicking human behavior, including realistic posting schedules~\cite{khaund2021social}, tweet content generation~\cite{grimme2022new}, and user profile imitation~\cite{heidari2020deep}. Moreover, by adaptively adjusting connection strategies~\cite{ghiuruau2024distinguishing}, bots can further obscure identities, making it increasingly difficult to distinguish them from genuine users.

\begin{figure}[t]
    \centering
    \includegraphics[width=.95\linewidth]{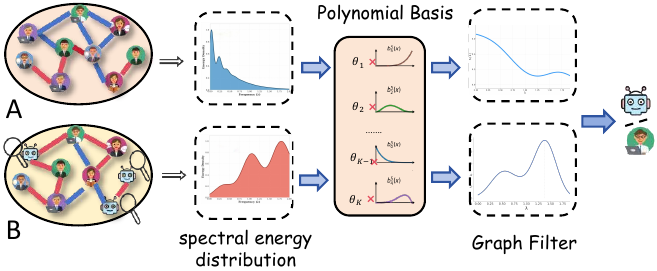}
    \caption{Illustration of homophily and spectral graph filtering in social bot detection. Homophilic community $A$: nodes with same labels form dense connections and concentrate spectral energy in low frequencies. Heterophilic community $B$: cross-label connections create dispersed patterns, shifting spectral energy toward higher frequencies. The spectral representations are then processed using polynomial basis functions to construct graph filters for bot detection.}
    \label{fig:homo&detection}
\end{figure}

Graph neural networks (GNNs) have become the mainstream approach for social bot detection by effectively leveraging both structural and attribute information~\cite{liu2023survey}. Existing GNN-based methods can be broadly divided into spatial and spectral approaches. Spatial methods focus on message passing and aggregation in the node domain~\cite{cai2021rethinking}, capturing local neighborhood information through iterative feature propagation. These methods typically employ graph convolution operations that aggregate information from neighboring nodes, enabling the model to learn node representations based on local structural context. While representative works such as BotRGCN~\cite{feng2021botrgcn}, BotMoE~\cite{liu2023botmoe}, and H2GCN~\cite{shao2024h2gcn} have made significant progress in this area and demonstrated strong performance in various social network scenarios, spatial approaches are inherently dependent on the topological structure of graphs, where different social networks feature distinct structural patterns and varying sparsity levels that may limit generalization~\cite{luan2023graph,huo2023t2}. 

Bot detection can be viewed as an anomaly detection problem~\cite{rafique2024machine}, making graph spectral methods particularly promising for this task. As illustrated in Figure~\ref{fig:homo&detection}, these methods leverage spectral graph theory and polynomial approximations to transform node features into the spectral domain, where polynomial-based filters are applied to capture global structural patterns and anomalies that may not be apparent in the spatial domain. A representative method is BWGNN~\cite{tang2022rethinking}, which uses Beta polynomial basis functions to capture spectral features. Through mathematical derivation and empirical visualization, BWGNN demonstrates the relationship between heterophily ratio and spectral energy distribution on graphs with varying heterophily levels, showing that higher heterophily leads to spectral energy concentration in higher frequency bands. Other notable spectral methods include BernNet~\cite{he2021bernnet}, JacobiConv~\cite{wang2022powerful}, and PolyGCL~\cite{chen2024polygcl}, achieving good performance in bot detection scenarios.

While spectral GNN methods are capable of fitting diverse spectral patterns through flexible polynomial filters~\cite{zeng2023graph,guo2023graph} and demonstrating strong performance in social bot detection tasks, they may be less sensitive to local bot-specific spectral anomalies that are crucial for effective bot detection. Most existing approaches use global polynomial filters, which fit the entire spectral domain by optimizing polynomial bases and their coefficients with respect to a loss defined over the full spectral domain. In practice, this global fitting may sometimes lose accuracy in critical bands where bot signals are most prominent, as anomalous behaviors tend to exhibit abrupt changes~\cite{yu2019unsupervised} that are not easily captured by filters optimized for overall spectral approximation.

In addition, such a valuable relationship between graph homophily and spectral energy has not been fully leveraged by the spectral GNN methods for improved bot detection performance. As illustrated in Figure~\ref{fig:homo&detection}, homophily characteristics directly influence spectral energy distribution~\cite{tang2022rethinking,gao2023addressing,xu2024revisiting}. This relationship provides valuable domain knowledge for determining spectral focus regions in bot detection, but is still underutilized in current spectral GNN approaches.
To address limitations above, we propose \methodname{}, a homophily-aware Gaussian-window constrained graph spectral network for bot detection. The contributions of our method are summarized as follows:

\begin{itemize}
\item \textbf{Gaussian-Window constrained Spectral Network:} We introduce a Gaussian‑window constrained spectral network that enables precise local spectral-domain fitting by employing learnable Gaussian functions as soft polynomial basis masks. This approach uses multiple Gaussian windows to extract features focusing on different frequency bands for comprehensive anomaly capture, providing precise local spectral filtering and plug‑in compatibility with existing spectral GNNs (e.g., BWGNN, BernNet).
\item \textbf{Homophily‑Aware Adaptation Mechanism:} We innovatively inject domain knowledge by measuring the graph's homophily ratio and using a frequency distribution loss to adaptively guide the Gaussian window parameters, ensuring that spectral focus aligns with homophily‑driven frequency preferences. This mechanism incorporates domain knowledge from bot detection to adaptively guide spectral focus.
\end{itemize}

Extensive experiments on five widely used real-world benchmarks demonstrate that \methodname{} achieves an average improvement of 4.3\% in F1-score over state-of-the-art baselines, validating the effectiveness and plug-in compatibility of our approach.

\begin{figure*}[htb]
  \centering
  \includegraphics[width=\textwidth]{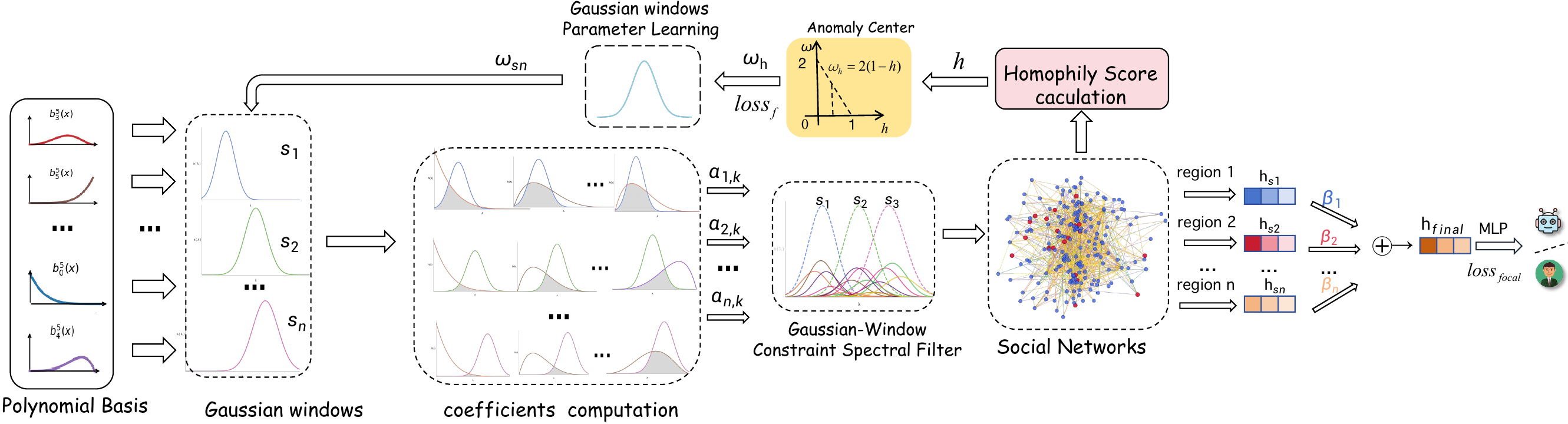}
    \caption{\methodname{} Framework: homophily-aware Gaussian-window constrained graph spectral network for social bot detection. The framework employs learnable Gaussian windows to modulate the weights of polynomial basis functions, enabling focused spectral analysis on bot-discriminative frequency bands. The homophily-aware adaptation mechanism injects domain knowledge between homophily ratios and frequency features to guide window parameter learning.}
  \label{fig:framework}
\end{figure*}
\section{Related Work}
In recent years, graph-based approaches have emerged as a promising direction for social bot detection, as such approaches enable the integration of relational and user profile information within social networks. In this section, we review related work on graph-based bot detection, focusing on GNNs for homophilic and heterophilic graphs, spatial GNN methods, and spectral GNN methods.

Early GNN models such as GCN~\cite{kipf2016semi}, GraphSAGE~\cite{hamilton2017inductive}, and SGC~\cite{wu2019simplifying} are primarily designed under the homophily assumption, where nodes with similar labels are densely connected. However, many real-world social networks exhibit heterophilic or mixed connectivity patterns, where nodes with different labels are frequently connected. To address this challenge, a series of methods have been developed to better handle heterophilic graphs, such as CPGNN~\cite{zhu2021graph}, FAGCN~\cite{bi2022feature}, and H2GCN~\cite{shao2024h2gcn}. These approaches relax the homophily assumption and enable more effective information aggregation in heterophilic settings, improving the robustness and generalization of GNNs in complex scenarios.

Spatial GNN methods leverage network topology and neighborhood aggregation to capture structural patterns, achieving strong performance in bot detection. Meanwhile, methods such as RGT~\cite{feng2022heterogeneity} and SlimG~\cite{yoo2023less} further improve robustness and interpretability. For bot detection, specialized spatial GNNs have been proposed, including BotRGCN~\cite{feng2021botrgcn}, BotMoE~\cite{liu2023botmoe}, and BSG4Bot~\cite{miao2025bsg4bot}, which introduce relational modeling, mixture-of-experts, or subgraph construction to better capture diverse bot behaviors. While spatial GNNs are effective at modeling local patterns, their effectiveness is constrained by the graph topology, as diverse social networks exhibit varying structural characteristics and connectivity densities that restrict their adaptability~\cite{luan2023graph,tang2020investigating}.

Spectral GNN methods operate in the spectral domain to capture global frequency-specific patterns. Representative approaches include ChebNet~\cite{defferrard2016convolutional}, BernNet~\cite{he2021bernnet}, GPR-GNN~\cite{chien2020adaptive}, BWGNN~\cite{tang2022rethinking}, and JacobiConv~\cite{wang2022powerful}, which utilize various polynomial bases for spectral filtering. More recent works such as PolyGCL~\cite{chen2024polygcl} that combines polynomial filters with self-supervised learning, and TFGNN~\cite{li2025polynomial} further enhance the expressiveness and adaptability of spectral filters. While these methods can fit a wide range of frequency responses, most existing spectral GNNs may lack sensitivity to local spectral anomalies and often do not explicitly incorporate domain knowledge, which is crucial for distinguishing bots from genuine users in complex networks.

Different from prior work, \methodname{} introduces learnable Gaussian windows to locally focus spectral filtering on critical frequency bands. Furthermore, it adapts Gaussian window's parameters based on the graph's homophily ratio to leverage bot detection domain knowledge, ensuring precise anomaly capture aligned with structural homophily.

\section{Methodology}
The framework of \methodname{} is shown in Figure~\ref{fig:framework}. It consists of two major components: Gaussian-Window constrained Spectral Network and Homophily-Aware Adaptation Mechanism. First, the Gaussian-Window constrained Spectral Network employs learnable Gaussian windows with polynomial approximation to concentrate filtering on discriminative frequency bands, enabling precise spectral analysis for bot detection. Subsequently, the Homophily-Aware Adaptation Mechanism measures graph homophily ratios and adaptively tunes window parameters to ensure spectral focus aligns with anomaly characteristics, achieving superior performance through valuable domain knowledge injection.

\subsection{Problem Formulation}

We represent the social network as a heterogeneous attributed graph \(G=(V,E,\mathcal{X},\mathcal{R})\), where \(V\) is the set of users, \(\mathcal{X}\in\mathbb{R}^{n\times d_0}\) their feature vectors, and \(\mathcal{R}\) the relation types (e.g., “follows,” “mentions,” “retweets”) with edges \(E_r\subseteq V\times V\). While our framework supports heterogeneous graphs and can integrate information from different relation types, we focus on single-relation GNNs in the following, and multiple relationships can be supported through attention-based methods. A subset \(V_L\subseteq V\) has binary labels \(y_i\in\{0,1\}\) indicating genuine and bot users. We aim to learn
\begin{equation}
\label{eq:graph_mapping}
f: G \longmapsto [0,1]^n
\end{equation}
to predict each node’s bot probability \(\hat y_i\).

To adapt to mixed homophily, we compute the graph-level homophily ratio~\cite{ma2021homophily}, and assume that the target frequency band is inversely correlated with the homophily ratio, \emph{i.e.}, lower homophily leads to a preference for higher spectral components.
\begin{equation}
\label{eq:homophily_ratio}
h = \frac{\bigl|\{(u,v)\in E : y_u = y_v\}\bigr|}{|E|}
\end{equation}
where \(E=\bigcup_{r}E_r\). The scalar \(h\) can be used to guide the positioning of Gaussian windows in our spectral filters.  

\subsection{Gaussian-Window constrained Spectral Network}

The core innovation of our approach lies in introducing learnable Gaussian windows that modulate polynomial basis coefficients to achieve local spectral focusing on bot-discriminative frequency bands, enabling the model to extract and combine the most informative spectral features for distinguishing bots from genuine users.

\subsubsection{Gaussian Window Spectral Focusing Strategy.}

While spectral GNNs with global filters can capture various spectral patterns, they may degrade focus to detect abrupt frequency changes associated with anomaly features. To address this issue, we employ $S$ learnable Gaussian windows that focus on specific frequency bands for comprehensive bot detection:

\begin{equation}
\label{eq:gaussian_filter}
G_s(\lambda) = \exp\!\left(-\tfrac{(\lambda - \omega_s)^2}{2\sigma_s^2}\right),\quad s = 1,\dots,S
\end{equation}

Here \(\lambda\in[0,2]\) is a Laplacian eigenvalue, \(\omega_s\) the center frequency, and \(\sigma_s\) the window's bandwidth. Each Gaussian window $G_s(\lambda)$ defines a localized spectral filter that focuses on specific frequency regions. By learning \(\{\omega_s,\sigma_s\}\), the model adaptively adjusts the center frequencies and bandwidths of Gaussian windows to capture information from different spectral regions and the combination of multiple such filters enables comprehensive spectral analysis.

Direct spectral filtering requires eigendecomposition of the Laplacian matrix with computational complexity $\mathcal{O}(n^3)$, which is prohibitive for large graphs. 
Therefore, for our Gaussian-window constrained filters $g_s(\lambda)$, we use polynomial basis functions $\{P_k^{(K)}(\lambda)\}_{k=0}^K$ of order $K$ to approximate the filter response:
\begin{equation}
\label{eq:polynomial_filter_approx}
g_s(\lambda) \approx \sum_{k=0}^{K} c_{s,k} P_k^{(K)}(\lambda)
\end{equation}

The interaction coefficient $c_{s,k}$ reflects how much each polynomial basis function contributes to approximating the $s$-th Gaussian-Window constrained filter. For all polynomial basis functions $\{P_k^{(K)}(\lambda)\}_{k=0}^K$ and a given Gaussian window $G_s(\lambda)$, we compute the contribution weight of each basis function:
\begin{equation}
\label{eq:interaction_coeff}
c_{s,k} = \int_0^2 G_s(\lambda) P_k^{(K)}(\lambda) \, d\lambda
\end{equation}
These coefficients are determined by the overlap between the Gaussian window and the polynomial basis functions in spectral region. \textit{The \textbf{greater} the overlap between a polynomial basis and the Gaussian window's frequency region, the \textbf{larger} the corresponding coefficient $c_{s,k}$, resulting in stronger emphasis on the corresponding spectral region.}

\subsubsection{Gaussian-Window constrained Spectral Filter Implementation.}

With the coefficients obtained from the Gaussian window analysis, we implement the spectral filtering by integrating multiple Gaussian-windowed filters. 

Starting from the spectral filter with integration coefficients, we derive the complete implementation through the following steps:
\begin{align}
g_s(\mathbf{L}) &= \int_0^2 G_s(\lambda) \, d\mathbf{E}(\lambda) \label{eq:spectral_integral}\\
&\approx \sum_{k=0}^K c_{s,k} P_k^{(K)}(\mathbf{L}) \label{eq:polynomial_approx_matrix}\\
\mathbf{Z}_s &= g_s(\mathbf{L}) \mathbf{X} \mathbf{W}_s \label{eq:window_filtering}\\
&= \left(\sum_{k=0}^K c_{s,k} P_k^{(K)}(\mathbf{L})\right) \mathbf{X} \mathbf{W}_s \label{eq:expand_filtering}\\
&= \sum_{k=0}^K c_{s,k} P_k^{(K)}(\mathbf{L}) \mathbf{X} \mathbf{W}_s \label{eq:final_window_conv}
\end{align}
where \(\mathbf{E}(\lambda)\) is the spectral measure, \(c_{s,k}\) are the coefficients from Eq.~\eqref{eq:interaction_coeff} that reflect the overlap between Gaussian windows and polynomial basis functions, \(P_k^{(K)}(\mathbf{L})\) are the polynomial basis matrices applied to the Laplacian, \(\mathbf{X}\) is the input feature matrix, and \(\mathbf{W}_s \in \mathbb{R}^{d \times d'}\) is the learnable weight matrix for the \(s\)-th window. 

The multi-window output combines all spectral features through learned weights:
\begin{align}
\mathbf{H} &= \sum_{s=1}^S w_s \mathbf{Z}_s \label{eq:multi_window_combine}\\
&= \sum_{s=1}^S w_s \sum_{k=0}^K c_{s,k} P_k^{(K)}(\mathbf{L}) \mathbf{X} \mathbf{W}_s \label{eq:complete_conv}
\end{align}

where \(w_s\) are learned importance weights computed through a softmax function:
\begin{equation}
\label{eq:window_weights}
w_s = \frac{\exp(\beta_s)}{\sum_{j=1}^S \exp(\beta_j)}
\end{equation}
with learnable parameters \(\beta_s\), ensuring \(\sum_{s=1}^S w_s = 1\).


\subsection{Homophily-Aware Adaptation Mechanism}

The center frequencies and bandwidths of Gaussian windows should adapt to the structural characteristics of different social networks. Our solution leverages the relationship between graph homophily and spectral energy distribution to guide the learning of optimal window parameters.




\subsubsection{Gaussian Window Parameter Learning.}

For each edge type \(e\) with homophily ratio \(h_e\), we set the expected anomaly center frequency as
\begin{equation}
\label{eq:omega_target}
\bar\omega(h_e) = 2\,(1 - h_e)
\end{equation}
This mapping reflects the spectral theory insight that high homophily (\(h_e \to 1\)) concentrates energy in low frequencies (\(\bar\omega \to 0\)), while low homophily (\(h_e \to 0\)) shifts energy toward high frequencies (\(\bar\omega \to 2\)).

While the above relationship provides a theoretical mapping between homophily ratios and center frequencies, this mapping is empirical and should not strictly constrain the learned frequencies. We allow some deviation around the target frequencies to provide flexibility in learning. To inject this homophily-driven domain knowledge into the spectral filtering process, we employ MLP networks to learn the center frequencies and bandwidths of the Gaussian windows. Specifically, the MLPs take the homophily-guided target frequency $\bar{\omega}(h_e)$ as input and output the parameters for each window:
\begin{align}
\omega_s &= \text{MLP}_{\omega}(\bar{\omega}(h_e), s) \label{eq:omega_learn}\\
\sigma_s &= \text{MLP}_{\sigma}(\bar{\omega}(h_e), s) \label{eq:sigma_learn}
\end{align}
where \(\text{MLP}_{\omega}: \mathbb{R}^2 \to [{s}^{-}, {s}^{+}]\) and \(\text{MLP}_{\sigma}: \mathbb{R}^2 \to \mathbb{R}^+\) are multi-layer perceptrons. The center frequencies \(\omega_s\) are initialized as \(S\) equally-spaced points in the spectral domain \([0, 2]\), which provides comprehensive coverage of the entire frequency range. During training, the MLPs learn adaptive offsets from these initial positions toward the homophily-guided target \(\bar{\omega}(h_e)\), enabling the model to adjust spectral focus based on graph structural properties. To maintain local spectral focus and prevent excessive deviation, the MLP output is clipped to a constrained range \([{s}^{-}, {s}^{+}]\) around each window's initial position, ensuring that each Gaussian window remains concentrated on its designated local frequency region. By explicitly feeding the homophily-guided target frequency into the MLPs, we ensure that the parameterization of each Gaussian window is directly influenced by the structural properties of the graph. Furthermore, the entire process is trained end-to-end with a frequency distribution loss, which continuously encourages the MLPs to produce window parameters that remain closely aligned with the homophily-driven spectral targets. This design enables the model to flexibly adapt to diverse network structures while still preserving the theoretical connection between homophily and spectral focus.

To further reinforce this alignment, we introduce a frequency distribution loss that anchors the learned window centers to positions determined by the homophily ratio:
\begin{equation}
\label{eq:freq_loss}
\mathcal{L}_{\mathrm{freq}}
= \frac{1}{C}\sum_{c=1}^C
  \bigl(\hat\omega^{(c)} - \bar\omega(h_e)\bigr)^2
\end{equation}
where \(\hat\omega^{(c)}\) is the learned center frequency in the convolution block $c$ and $C$ is the total number of blocks.

The Gaussian window parameter learning mechanism thus injects domain knowledge by measuring the graph's homophily ratio and uses a frequency distribution loss to guide the Gaussian window parameters, ensuring that spectral focus aligns with homophily-driven frequency preferences.





\begin{table*}[!t]
\centering
\caption{Accuracy and F1-score of Competitors on the Five Benchmarks.}
\label{tab:performance}
{\small
\begin{tabular}{@{}lcccccccccc@{}}
\toprule
Model        & \multicolumn{2}{c}{Twibot-20} & \multicolumn{2}{c}{Twibot-22} & \multicolumn{2}{c}{MGTAB} & \multicolumn{2}{c}{T-social} & \multicolumn{2}{c}{T-finance}\\
             & Accuracy       & F1     & Accuracy      & F1      & Accuracy      & F1  & Accuracy      & F1   & Accuracy      & F1\\
\midrule
MLP         &83.89  &81.71   &79.01   &53.81   &84.88     &84.67    &71.52   &48.35       &78.85   &70.57   \\
\midrule
GCN         &77.52     &80.85     &78.41      &54.91     &83.65     &84.02   &75.87   &59.88       &79.03   &70.74    \\
GAT         &83.33   &81.26     &\underline{79.54}    &55.83    &84.45     &83.69  &76.82    &69.01       &77.21   &53.86   \\
BotRGCN     &85.86   &87.33     &78.56     &57.52    &89.69     &86.02   &77.65   & 68.13    &75.98   & 52.14   \\
H2GCN        &88.23  &89.14  &77.64  &57.23  &90.56  &87.72   &--   &--       &87.93   &88.24    \\
GPR-GNN      &87.47 &88.84 &78.64 &57.66      &90.32   &87.46  &75.13   &59.76     &84.78   &85.07   \\
SlimG       &86.55   &87.97     &74.76     &44.27    &88.13    &84.45 &74.76  &58.73     &82.21   &84.54   \\
RGT         &86.67   &88.22     &76.44    &43.02     &89.76   &86.59  &77.98  &71.23 &86.27   &87.12   \\
BSG4Bot   &\underline{89.15}  &\underline{89.89} & 79.53 & 59.42 &\underline{91.75} &88.53   &68.21   &66.15       &87.23   &86.82    \\
\midrule
BWGNN       &85.23   &87.62   &58.21    &59.78  &89.54    & 87.13  &81.43   &83.98     &85.76   &86.87    \\
BernNet     &87.78   &88.43   &78.93     & 59.98    &79.56  & 88.43  &76.56  & 84.13    &86.16   & 87.63   \\
JacobiConv  &84.58    &88.01   &59.94     &60.03  &90.72    &87.96   &61.19   &75.51  &82.21   &86.91    \\
TFGNN   &88.98    &89.13   &59.89    &59.71  &89.98    &88.14  &81.29  &84.07  &87.89   &88.05    \\
\midrule
HW-Beta  &87.23  &88.16  &77.54  &\underline{60.98}  &91.43  &88.93 &89.50 &91.16 &88.16 &89.59\\
HW-Jacobi  &88.80  &88.50  &78.84  &60.31  &89.45  &\textbf{89.87} &\textbf{96.71} &\textbf{94.47} &\textbf{90.46} &\underline{91.67}\\
HW-BernStein  &\textbf{90.37}&\textbf{91.51}&\textbf{80.73}&\textbf{61.95}&\textbf{92.83} &\underline{89.37} &\underline{96.53} &\underline{94.39} &\underline{88.91} &\textbf{91.95}\\
\midrule
\textit{Improvement}&\textbf{+1.3\%}&\textbf{+1.8\%}&\textbf{+1.4\%}&\textbf{+3.2\%}&\textbf{+1.2\%} &\textbf{+1.1\%}&\textbf{+18.5\%} &\textbf{+12.2\%} &\textbf{+3.2\%} &\textbf{+4.2\%}\\
\bottomrule
\end{tabular}
}
\end{table*}





\subsection{Multi-Layer Architecture}

We stack multiple Gaussian-windowed convolution layers with residual connections:
\begin{equation}
\label{eq:multilayer}
\mathbf{H}^{(\ell+1)} = \sigma\left(\mathbf{H}^{(\ell)} + \text{HW-Conv}(\mathbf{H}^{(\ell)})\right)
\end{equation}

where \(\mathbf{H}^{(\ell)}\) denotes features at layer \(\ell\), \(\sigma(\cdot)\) is the activation function, HW-Conv denotes the Gaussian-window constrained spectral network, and the residual connection helps mitigate over-smoothing.

\subsection{Overall Loss Function}

The complete loss function incorporates the classification loss and frequency distribution loss:

\begin{equation}
\label{eq:total_loss}
\mathcal{L}
= \mathcal{L}_{\mathrm{focal}}
  + \lambda_f\,\mathcal{L}_{\mathrm{freq}}
\end{equation}

where \(\mathcal{L}_{\mathrm{focal}}\) is the Focal Loss for addressing class imbalance:
\begin{equation}
\label{eq:focal_loss}
\mathcal{L}_{\mathrm{focal}} = -\frac{1}{|V_L|}\sum_{i \in V_L} \alpha_i (1-\hat{y}_i)^{\gamma} \log(\hat{y}_i)
\end{equation}

where $|V_L|$ is the number of training nodes, $\hat{y}_i$ is the predicted probability for the true class, $\alpha_i$ is a balancing factor, and $\gamma$ is the focusing parameter. \(\lambda_f < 1\) is a weighting parameter that balances classification accuracy with frequency distribution regularization.


\subsection{Computational Complexity Analysis}
Let $S$ be the number of Gaussian windows, $K$ be the polynomial order, $|E|$ be the number of edges, $d$ and $d'$ be the input and output feature dimensions, and $d_{\text{MLP}}$ be the number of parameters in the MLPs for window parameter learning.

Our method consists of two main computational steps. For step (i), the Gaussian window parameter learning and polynomial coefficient computation requires $\mathcal{O}(S \cdot d_{\text{MLP}} + S \cdot K )$ operations. For step (ii), the multi-window filtering requires $\mathcal{O}(S \cdot K \cdot |E| \cdot d \cdot d')$ operations.

The overall computational complexity is $\mathcal{O}(S \cdot K \cdot |E| \cdot d \cdot d' + S \cdot d_{\text{MLP}} + S \cdot K)$. The dominant term $S \cdot K \cdot |E| \cdot d \cdot d'$ scales linearly with the graph size and feature dimensions. While our method introduces additional complexity for Gaussian window parameter learning compared to baseline polynomial methods, this overhead only affects the less computationally intensive coefficient computation step, ensuring scalability for large-scale social networks.

\section{Experiments}


\begin{figure*}[t]
  \centering

  \begin{subfigure}[t]{0.24\textwidth}
    \centering
    \includegraphics[width=\linewidth]{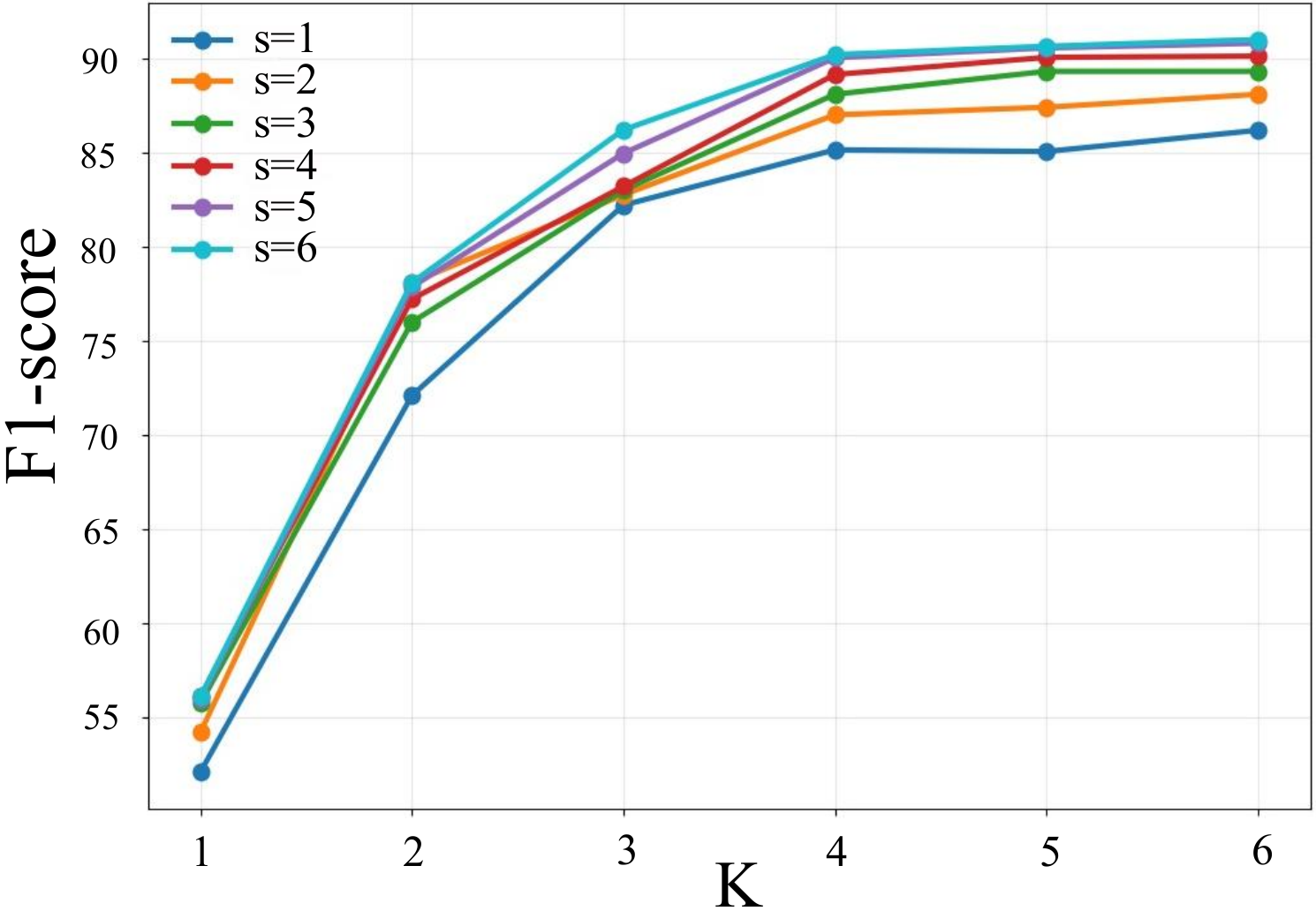}
    \caption{Twibot-20}
  \end{subfigure}
  \hfill
  \begin{subfigure}[t]{0.24\textwidth}
    \centering
    \includegraphics[width=\linewidth]{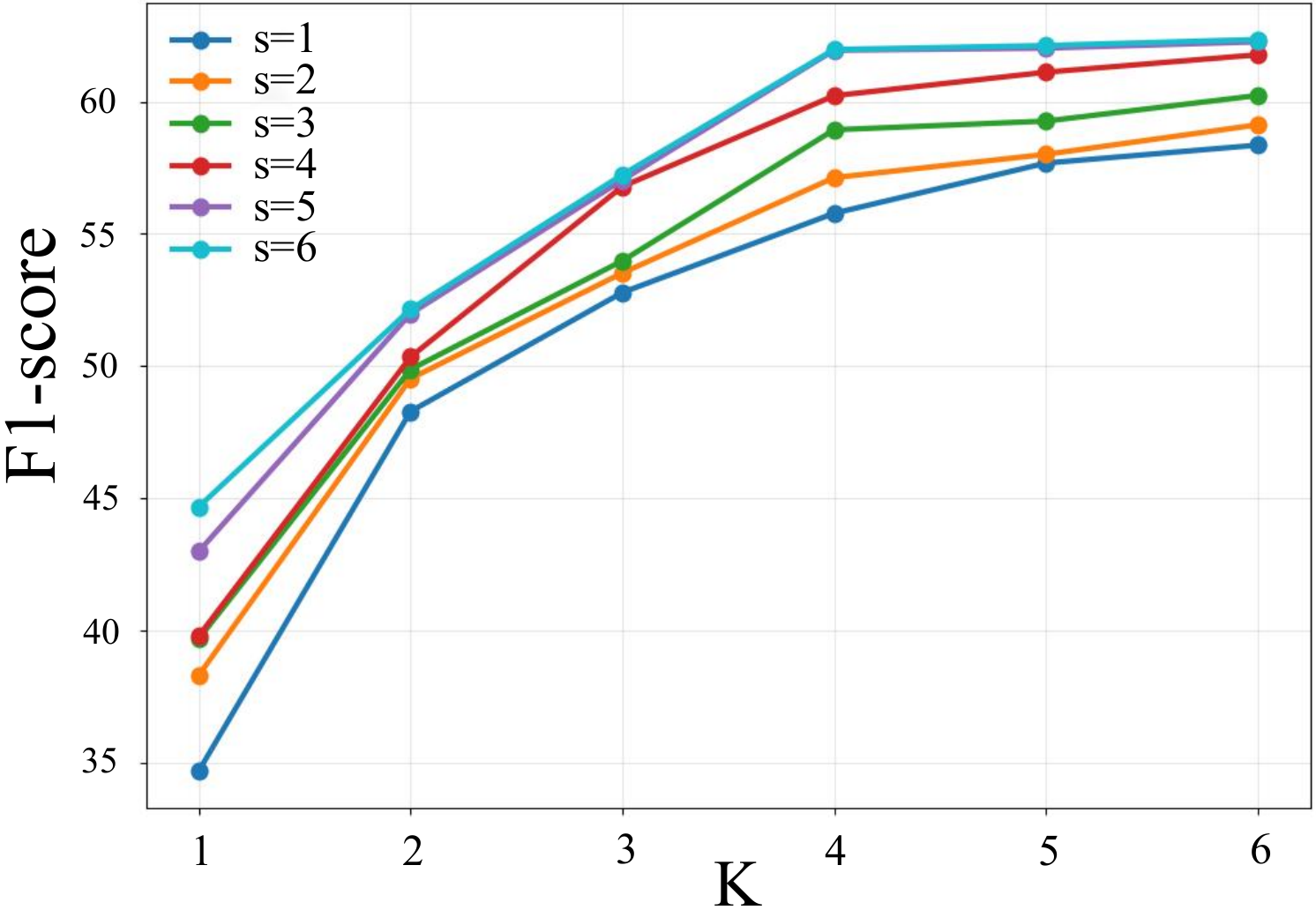}
    \caption{Twibot-22}
  \end{subfigure}
  \hfill
  \begin{subfigure}[t]{0.24\textwidth}
    \centering
    \includegraphics[width=\linewidth]{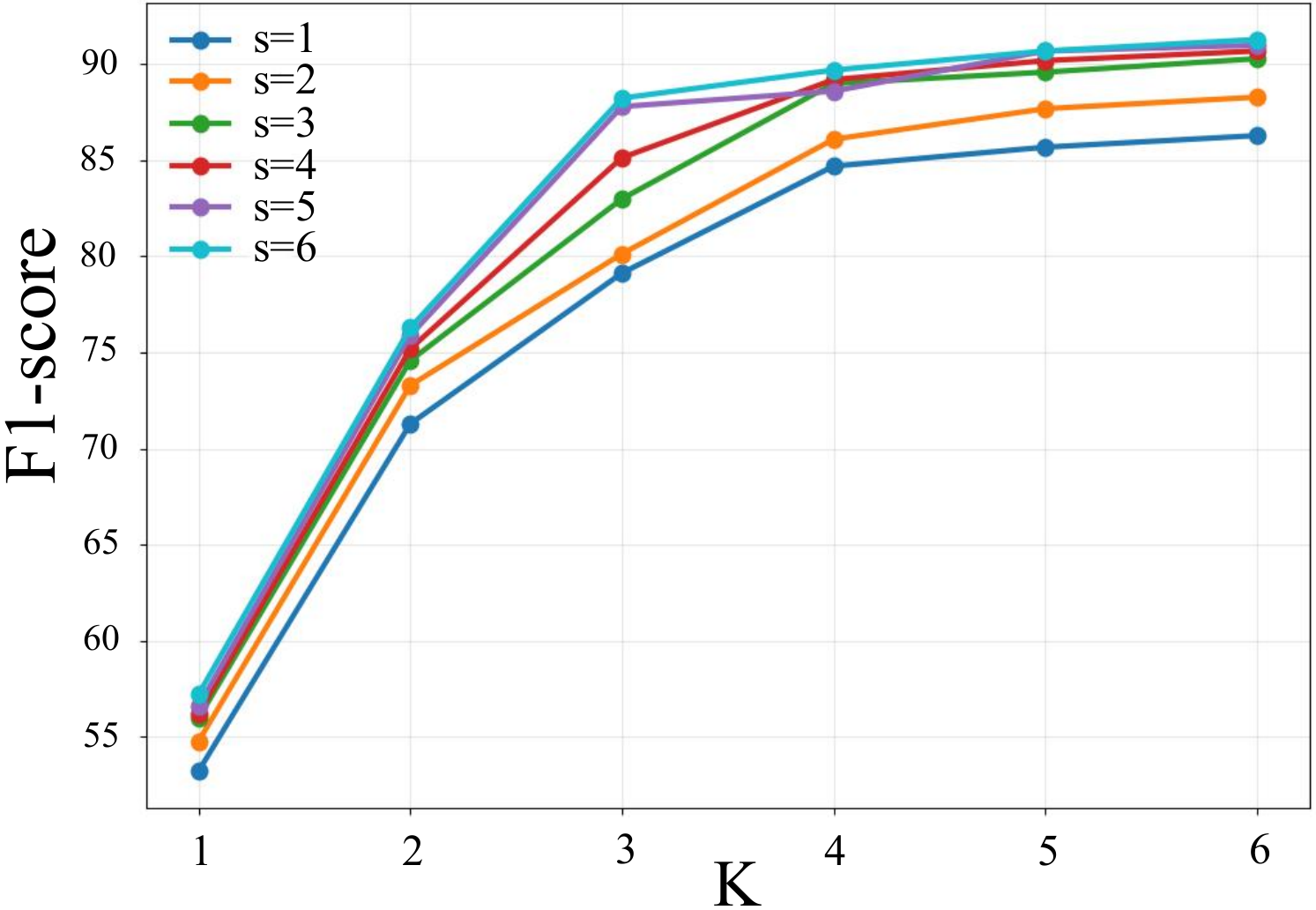}
    \caption{MGTAB}
  \end{subfigure}
  \hfill
  \begin{subfigure}[t]{0.24\textwidth}
    \centering
    \includegraphics[width=\linewidth]{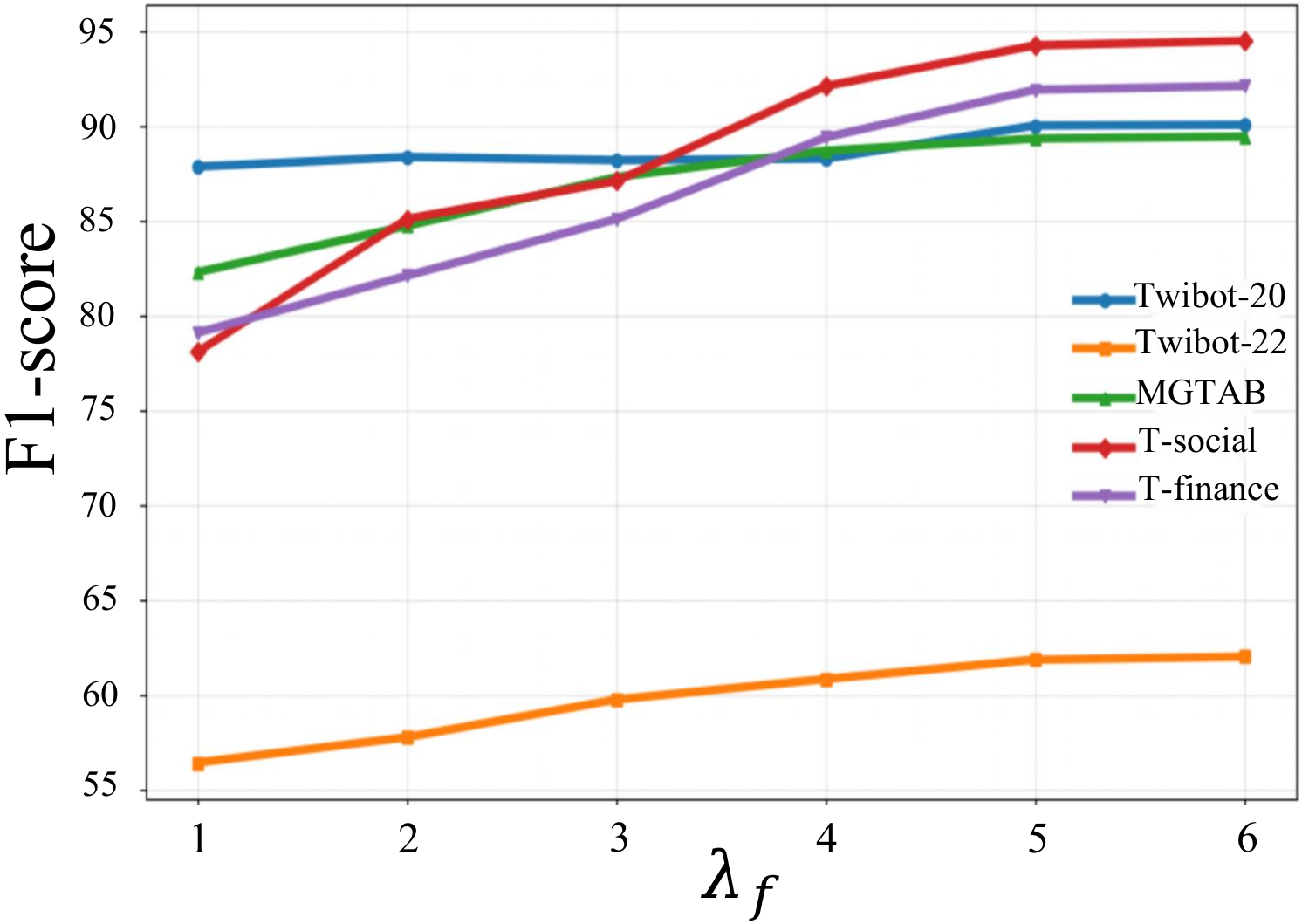}
    \caption{Loss weight $\lambda_f$}
  \end{subfigure}

  \caption{
    Sensitivity analysis of key parameters in \methodname{}. 
    (a)--(c): F1-score on Twibot20, Twibot22, and MGTAB datasets, respectively, with varying polynomial order $K$ and different numbers of Gaussian windows $S$ (each line represents a different $S$). 
    (d): F1-score variation with respect to the frequency distribution loss weight $\lambda_f$.
  }
  \label{fig:sensitivity_3}
\end{figure*}

\begin{figure*}[t]
  \centering

  \begin{subfigure}[t]{0.19\textwidth}
    \centering
    \includegraphics[width=\linewidth]{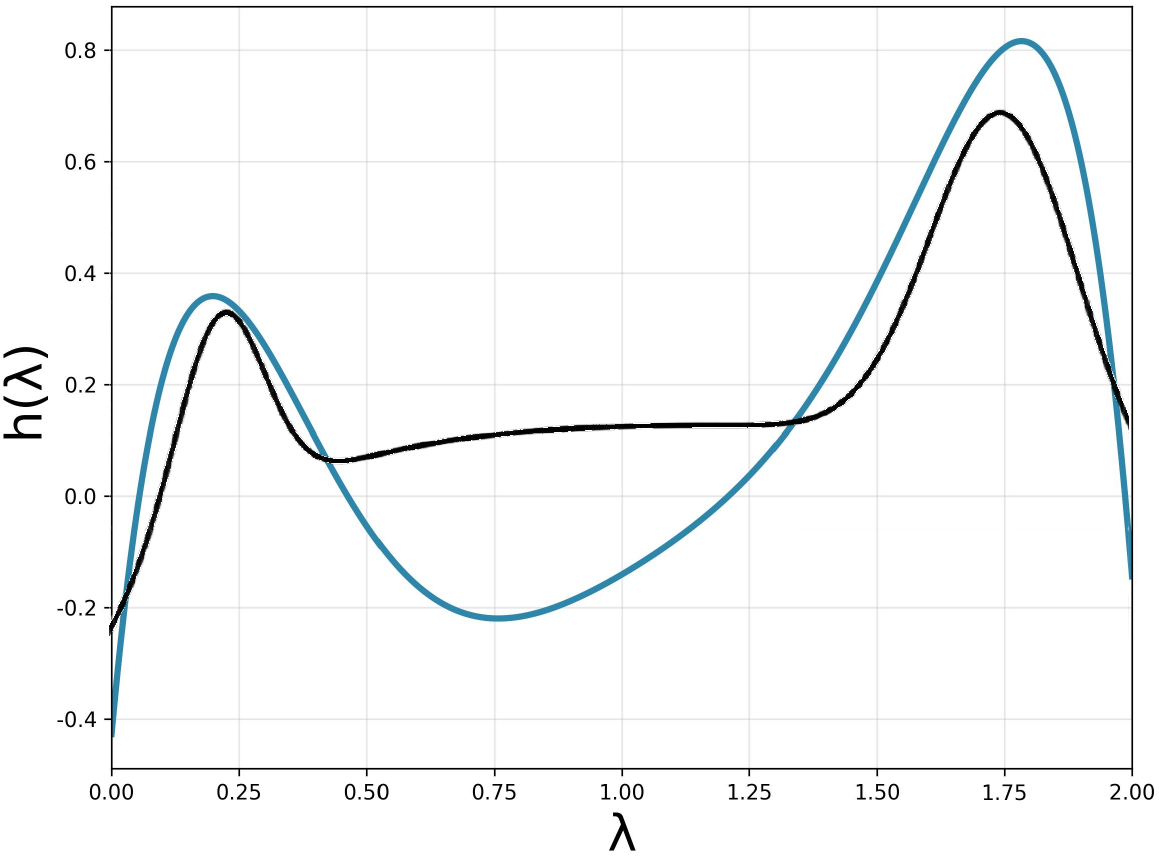}
    \caption{TwiBot-20}
  \end{subfigure}
  \hfill
  \begin{subfigure}[t]{0.19\textwidth}
    \centering
    \includegraphics[width=\linewidth]{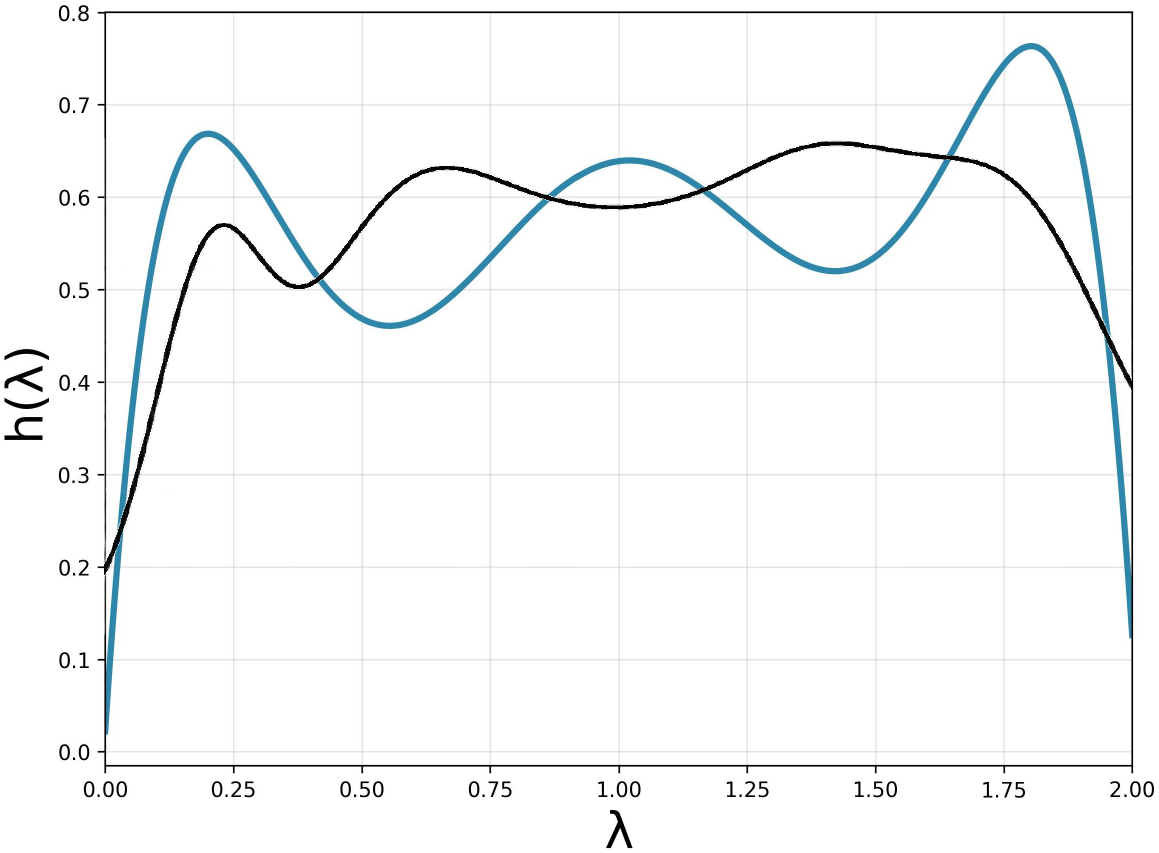}
    \caption{TwiBot-22}
  \end{subfigure}
  \hfill
  \begin{subfigure}[t]{0.19\textwidth}
    \centering
    \includegraphics[width=\linewidth]{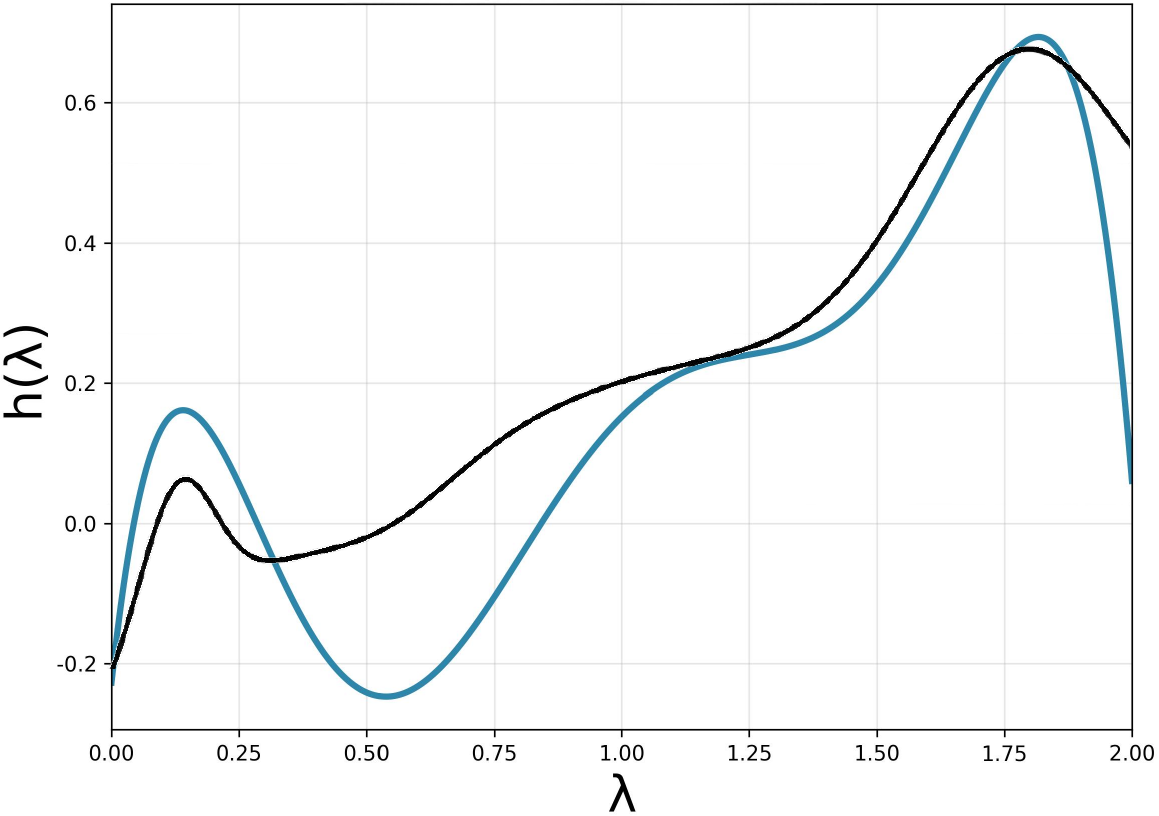}
    \caption{MGTAB}
  \end{subfigure}
  \hfill
  \begin{subfigure}[t]{0.19\textwidth}
    \centering
    \includegraphics[width=\linewidth]{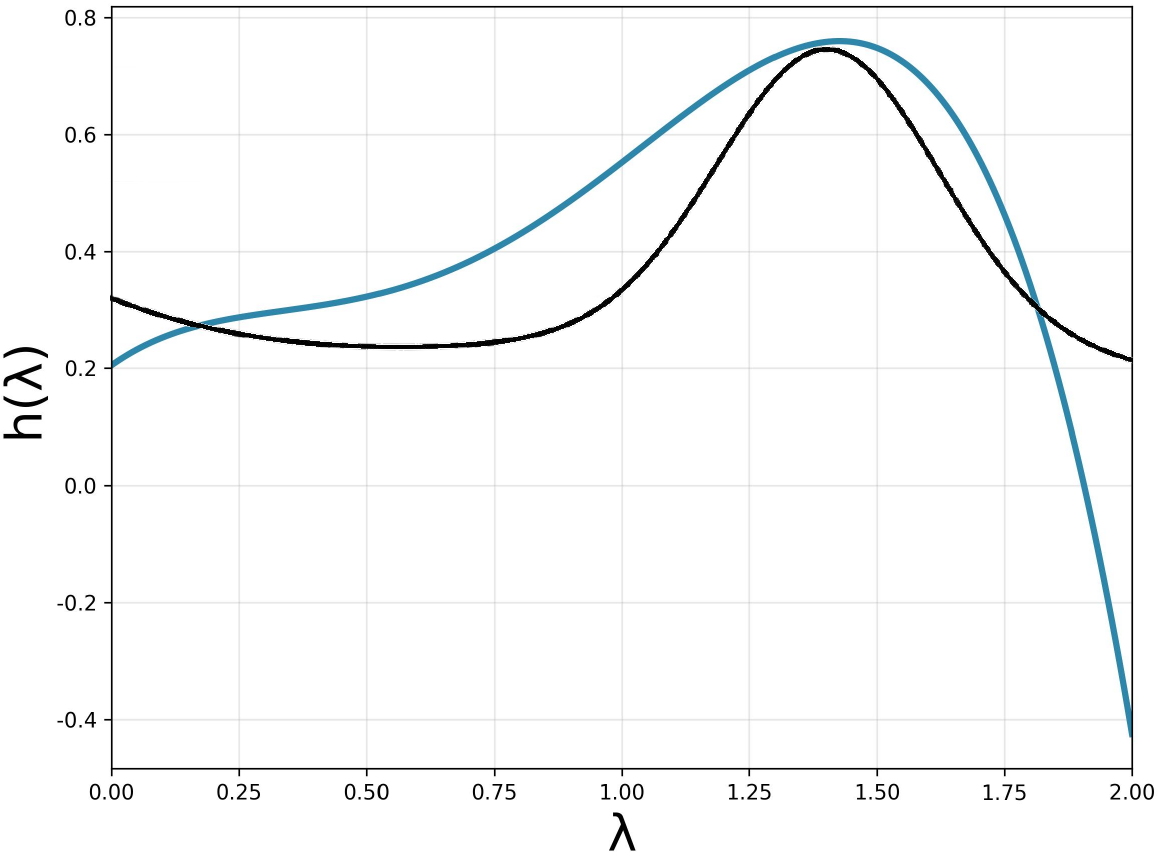}
    \caption{T-Finance}
  \end{subfigure}
  \hfill
  \begin{subfigure}[t]{0.19\textwidth}
    \centering
    \includegraphics[width=\linewidth]{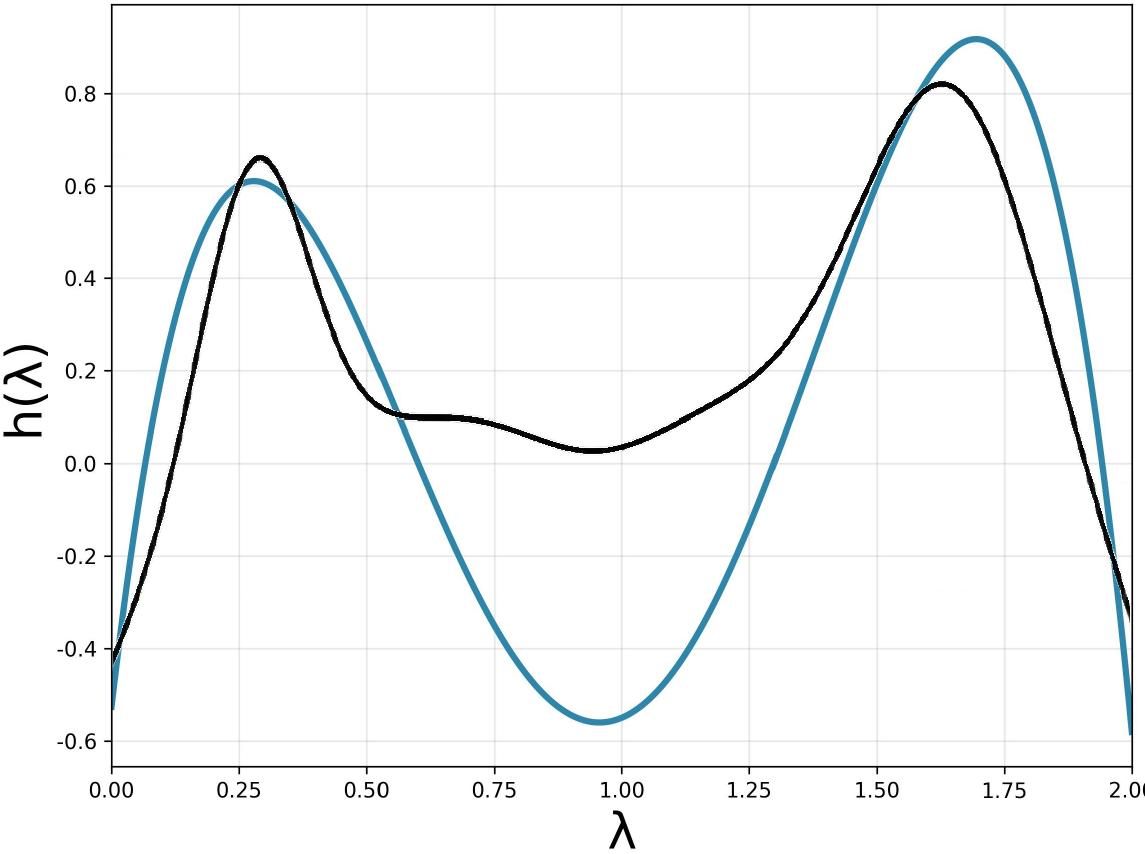}
    \caption{T-social}
  \end{subfigure}

  \caption{Learned spectral filter $h(\lambda)$ for different datasets by \methodname{} (blue) and original Bernstein polynomial (black).}
  \label{fig:spectral_filters}
\end{figure*}

\subsection{Experimental Setup}

\subsubsection{Datasets}

We evaluate \methodname{} on five widely‑adopted social bot detection benchmarks that span different scales, relation types, and homophily patterns. TwiBot‑20~\cite{feng2021twibot} comprises 229,580 Twitter users with follow and mention relations and a balanced mix of human and bot accounts. TwiBot‑22~\cite{feng2022twibot} scales up to 1,000,000 users under the same two relations, reflecting a more realistic, large‑scale detection scenario. MGTAB~\cite{shi2023mgtab} offers a heterogeneous, multi‑platform view with users connected by seven relation types. T‑Social~\cite{tang2022rethinking} represents a massive homophilic network of 5,781,065 users under a single follow relation, while T‑Finance~\cite{tang2022rethinking} focuses on financial conversations with 39,357 users and the same single relation. Detailed statistics of the datasets are provided in Appendix Section 2.1.

\subsubsection{Baselines}
We compare \methodname{} against representative baselines spanning three main categories. Content-based methods include MLP. Graph spatial methods encompass GCN~\cite{kipf2016semi}, GAT~\cite{velickovic2017graph}, BotRGCN~\cite{feng2021botrgcn}, RGT~\cite{feng2022heterogeneity}, SlimG~\cite{yoo2023less}, H2GCN~\cite{shao2024h2gcn}, and BSG4Bot~\cite{miao2025bsg4bot} which perform message passing and aggregation in the node spatial domain, focusing on local neighborhood information propagation. Graph spectral methods include GPR-GNN~\cite{chien2020adaptive}, BernNet~\cite{he2021bernnet}, BWGNN~\cite{tang2022rethinking}, JacobiConv~\cite{wang2022powerful}, and TFGNN~\cite{li2025polynomial} which leverage spectral graph theory and polynomial approximations to capture global graph properties through spectral domain analysis.

\begin{table*}[!t]
\centering
\caption{Ablation Study Results on the five Datasets.}
\label{tab:ablation}
{\small
\begin{tabular}{@{}lcccccccccc@{}}
\toprule
Model        & \multicolumn{2}{c}{Twibot-20} & \multicolumn{2}{c}{Twibot-22} & \multicolumn{2}{c}{MGTAB} & \multicolumn{2}{c}{T-social} & \multicolumn{2}{c}{T-finance}\\
             & Accuracy       & F1     & Accuracy      & F1      & Accuracy      & F1  & Accuracy      & F1   & Accuracy      & F1\\
\midrule
\methodname{} (Full) &\textbf{90.37}&\textbf{91.51}&\textbf{80.73}&\textbf{61.85}&\textbf{92.83}&\textbf{89.37}&\textbf{96.53}&\textbf{94.39}&\textbf{88.91}&\textbf{91.95}\\
\midrule
w/o Gaussian Window & 87.78 & 88.43 & 78.93 & 59.98  & 79.56 & 88.43 &76.56 & 84.13 &86.16 &87.63\\
w/o Homophily-Aware & 88.73 & 89.42 & 79.25 & 60.39 & 91.69 & 88.78  &94.59 &93.67 &88.98 &89.23\\
w/o Multi-band(S=1) &86.81  & 87.78 & 75.23 & 56.42 & 86.24 & 82.32 &92.83 &90.27 &86.95 &88.43\\
\bottomrule
\end{tabular}
}
\end{table*}

\subsubsection{Implementation Details}
\methodname{} is implemented in PyTorch and DGL. All experiments run on a server with 256GB RAM, dual Intel Xeon Silver CPUs @2.4GHz, and a NVIDIA RTXA6000 GPUs (48GB). To avoid overfitting, we use early stopping based on validation Macro‑F1. We set the number of Gaussian windows to 5 and polynomial order to 4 to achieve an optimal balance between accuracy and computational efficiency across all datasets.

\subsection{Performance on different Baselines}

The experimental results are presented in Table~\ref{tab:performance}. We compare \methodname{} against all baselines across the five benchmarks, reporting accuracy and F1-score results.

\methodname{} achieves superior performance across all benchmarks, demonstrating the effectiveness of our homophily-aware Gaussian-Window constrained spectral network. On TwiBot-20, our method outperforms the best baseline BSG4Bot by 1.2\% F1-score, showcasing the advantage of our Gaussian window mechanism in capturing bot-specific spectral features. For the challenging heterophilic TwiBot-22 dataset, we achieve 3.2\% F1 improvement over the second-best method JacobiConv, highlighting the effectiveness of our homophily-aware adaptation in handling complex heterophilic structures. Our method achieves the most significant improvements on large-scale datasets. T-Social achieves 18.5\% F1 improvement over BWGNN and T-Finance shows 4.2\% F1 improvement over BernNet, demonstrating the superior capability of our approach in processing large-scale social networks. On MGTAB, our method surpasses BSG4Bot by 1.2\% accuracy, further validating the robustness of our \methodname{} across diverse graph structures.

Notably, our \methodname{} demonstrates strong plug-in compatibility with existing spectral architectures and consistently outperforms their corresponding standalone polynomial GNNs. The results validate that our Gaussian-Window constrained spectral network can be effectively integrated with different polynomial bases (Beta, Jacobi, and Bernstein), with each HW variant showing significant improvements over the original polynomial methods, highlighting the broad applicability of our approach. The learned spectral filter responses $h(\lambda)$ comparing \methodname{} with original spectral methods are visualized in Figure~\ref{fig:spectral_filters}.

\subsection{Ablation Studies}
We perform ablation experiments using the Bernstein polynomial basis on five datasets to quantify the contribution of each component (Table~\ref{tab:ablation}). The results demonstrate that all components contribute significantly to the overall performance.
Removing the Gaussian-Window component causes F1 to drop ranging from 0.94\% to 10.26\%, confirming that the Gaussian-Window constrained spectral network is essential for effective spectral focusing, particularly in complex networks like T-social where the performance degradation is most pronounced. The homophily-aware adaptation mechanism also shows consistent importance, with F1 declines ranging from 0.59\% to 2.72\% when removed, validating the effectiveness of homophily domain knowledge injection for bot detection.

Moreover, replacing multiple Gaussian windows with a single global filter results in significant performance degradation, with F1 drops ranging from 2.28\% to 7.05\%. This demonstrates the critical necessity of multiple Gaussian windows for capturing diverse bot behavioral patterns across different frequency bands, particularly in networks where bots exhibit varied spectral features.
\subsection{Sensitivity Analysis of Key Parameters}

We study the sensitivity of \methodname{} to its main hyperparameters: number of Gaussian windows \(S\), polynomial order \(K\), and frequency distribution loss weight \(\lambda_f\).

As shown in Figure~\ref{fig:sensitivity_3}(a)-(c), model performance generally improves as both \(S\) and \(K\) increase. For all three datasets, the F1-score increases with \(K\) from 1 to approximately 4, then tends to plateau for higher \(K\) values. Higher \(S\) values consistently yield better performance than lower \(S\) values, with S=6 achieving the highest F1-scores across all datasets. The optimal performance is typically achieved within the range of \(S \in [4,6]\) and \(K \in [3,4]\), suggesting that \(S=5\) and \(K=4\) represent a practical balance between accuracy and efficiency.

Figure~\ref{fig:sensitivity_3}(d) illustrates the impact of the frequency distribution loss weight \(\lambda_f\) on model performance. The optimal \(\lambda_f\) varies across datasets. For TwiBot-22, MGTAB, and T-Finance, performance peaks at lower \(\lambda_f\) values (around \(\lambda_f \in [0.2,0.4]\)) and then tends to decrease with higher weights. Conversely, TwiBot-20 and T-social show a more robust or even slightly increasing trend at higher \(\lambda_f\) values after an initial peak. This indicates that the optimal strength of this loss weight depends on the specific characteristics of each dataset.

\section{Conclusion}

In this paper, we addressed the key limitation of existing spectral GNNs for their insufficient ability to achieve precise local region focusing and inject valuable homophily domain knowledge into bot detection. By employing Gaussian-window constrained spectral network and incorporating homophily-aware adaptation mechanism, \methodname{} achieves superior performance across diverse social bot detection benchmarks.

Future work will explore extending our homophily-aware mechanism to dynamic networks with evolving homophily patterns and integrating our approach with LLMs for enhanced social bot detection capabilities.

\bibliography{Reference}
\newcommand{\isChecklistMainFile}{}
\end{document}